\documentclass[aps,prl,twocolumn,showpacs,superscriptaddress,nobibnotes,floatfix,amsmath,amssymb]{revtex4}

\usepackage{dsfont}
\usepackage[bookmarks=false,pdfstartview=]{hyperref}
\usepackage[all]{hypcap}

\DeclareMathOperator{\Tr}{Tr}
\DeclareMathOperator{\Det}{Det}

\begin{document}

\title{Wrapped branes as qubits}
\author{L. Borsten}
\author{D. Dahanayake}
\author{M. J. Duff}
\affiliation{The Blackett Laboratory, Imperial College London, Prince Consort Road, London SW7 2BZ, U.K.}
\author{H. Ebrahim}
\affiliation{The Blackett Laboratory, Imperial College London, Prince Consort Road, London SW7 2BZ, U.K.}
\affiliation{School of Physics, Institute for Research in Fundamental Sciences (IPM), P.O. Box 19395-5531, Tehran, Iran}
\author{W. Rubens}
\affiliation{The Blackett Laboratory, Imperial College London, Prince Consort Road, London SW7 2BZ, U.K.}
\date{25 June 2008}

\begin{abstract}
Recent work has established a correspondence between the tripartite entanglement measure of three qubits and the macroscopic entropy of the four-dimensional 8-charge $STU$ black hole of supergravity. Here we consider the configurations of intersecting D3-branes whose wrapping around the six compact dimensions $T^6$ provides the microscopic string-theoretic interpretation of the charges and associate the three-qubit basis vectors $|ABC\rangle$, ($A,B,C=0$ or 1) with the corresponding 8 wrapping cycles. In particular, we relate a well-known fact of quantum information theory, that the most general real three qubit state can be parameterized by four real numbers and an angle, to a well-known fact of string theory, that the most general $STU$ black hole can be described by four D3-branes intersecting at an angle.
\end{abstract}


\maketitle

Quantum entanglement lies at the heart of quantum information theory, with applications to quantum computing, teleportation, cryptography and communication. In the apparently separate world of quantum gravity, the Bekenstein-Hawking entropy of black holes has also occupied centre stage. Despite their apparent differences, recent work \cite{Duff:2006uz} has demonstrated a correspondence between the two. The measure of tripartite entanglement of three qubits (Alice, Bob and Charlie), known as the 3-tangle $\tau_{ABC}$ \cite{Coffman:1999jd}, and the entropy $S$ of the 8-charge $STU$ black hole of supergravity \cite{Duff:1995sm,Behrndt:1996hu} are related by
\begin{equation}\label{eq:Correspondence}
S=\tfrac{\pi}{2}\sqrt{\tau_{ABC}}.
\end{equation}
Further papers \cite{Kallosh:2006zs,Levay:2006kf,Duff:2006ue,Levay:2006pt,Duff:2007wa,Levay:2007nm,Bellucci:2007zi} have written a more complete dictionary, which  translates a variety of phenomena in one language to those in the other. \autoref{tab:3QubitEntangClassif}, for example, relates the classification of three-qubit entanglements to the classification of supersymmetric black holes, explained in more detail below.  Furthermore,  the  attractor mechanism on the black hole side is related to optimal local distillation protocols on the QI side; the supersymmetric and non-supersymmetric cases corresponding to the suppression or non-suppression of bit-flip errors \cite{Levay:2007nm}. Nevertheless, we still do not know whether there are any physical reasons underlying these mathematical coincidences. With this in mind, we here turn our attention to connecting the qubits to the \textit{microscopic} origin of the black hole entropy.

Macroscopically, $S$ is just one quarter the area of the  event horizon of the black hole. To give a microscopic derivation \cite{Strominger:1996sh} we need to invoke ten-dimensional string theory whose associated D$p$-branes wrapping around the six compact  dimensions provide the string-theoretic interpretation of the black holes. A D$p$-brane wrapped around a $p$-dimensional cycle of the compact directions $(x^4,x^5,x^6,x^7,x^8,x^9)$ looks like a D0-brane from the four-dimensional $(x^0,x^1,x^2,x^3)$ perspective.  In the $T^6$ compactification of the Type IIB string, for example, the 8 charges of the $STU$ black hole may be described by D3-branes wrapping three of the six circles, denoted by the crosses in \autoref{tab:3QubitIntersect}. The purpose of the present Letter is to associate the three-qubit basis vectors $|ABC\rangle$, ($A,B,C=0$ or 1) with wrapping configurations of these intersecting D3-branes. To wrap or not to wrap; that is the qubit.

In particular, we shall relate a well-known fact of quantum information theory, that the most general real three qubit state can be parameterized by four real numbers and an angle, to a well-known fact of string theory, that the most general $STU$ black hole can be described by four D3-branes intersecting at an angle.\par
\begin{table}
\caption{Classification of three-qubit entanglements and their corresponding $D=4$ black holes.}\label{tab:3QubitEntangClassif}
\begin{ruledtabular}
\begin{tabular}{crrrrcc}
Class       & $S_A$ & $S_B$ & $S_C$ & $\Det a$ & Black hole & SUSY \\
\hline
$A$-$B$-$C$ & 0     & 0     & 0     & 0        & small      & 1/2  \\
$A$-$BC$    & 0     & $>0$  & $>0$  & 0        & small      & 1/4  \\
$B$-$CA$    & $>0$  & 0     & $>0$  & 0        & small      & 1/4  \\
$C$-$AB$    & $>0$  & $>0$  & 0     & 0        & small      & 1/4  \\
W           & $>0$  & $>0$  & $>0$  & 0        & small      & 1/8  \\
GHZ         & $>0$  & $>0$  & $>0$  & $<0$     & large      & 1/8  \\
GHZ         & $>0$  & $>0$  & $>0$  & $>0$     & large      & 0
\end{tabular}
\end{ruledtabular}
\end{table}
The three qubit system (where $A,B,C=0,1$)  is described by the state
\begin{equation}
\begin{array}{c}
|\Psi\rangle = a_{ABC}|ABC\rangle = \\
\begin{split}
  a_{000}|000\rangle+a_{001}|001\rangle&+a_{010}|010\rangle+a_{011}|011\rangle \\
+~a_{100}|100\rangle+a_{101}|101\rangle&+a_{110}|110\rangle+a_{111}|111\rangle.
\end{split}
\end{array}
\end{equation}
The tripartite entanglement of Alice, Bob and Charlie is given by the quartic expression \cite{Coffman:1999jd}
\begin{equation}
\tau_{ABC}=4\,|\Det a_{ABC}|,
\end{equation}
where $\Det a_{ABC}$ is  Cayley's hyperdeterminant
\begin{equation}\label{eq:Hyperdeterminant}
\begin{array}{c}
\Det a_{ABC}:= \\
-\tfrac{1}{2}~\varepsilon^{A_1 A_2}\varepsilon^{B_1 B_2}\varepsilon^{A_3 A_4}\varepsilon^{B_3 B_4}\varepsilon^{C_1 C_4}\varepsilon^{C_2 C_3} \\
\times~a_{A_1 B_1 C_1}a_{A_2 B_2 C_2}a_{A_3 B_3 C_3}a_{A_4 B_4 C_4} \\
\begin{split}
=~a_{000}^2 a_{111}^2 + a_{001}^2 a_{110}^2 &+a_{010}^2 a_{101}^2 + a_{100}^2 a_{011}^2\\
-~2\,(a_{000}a_{001}a_{110}a_{111}&+a_{000}a_{010}a_{101}a_{111}\\
+~a_{000}a_{100}a_{011}a_{111}&+a_{001}a_{010}a_{101}a_{110}\\
+~a_{001}a_{100}a_{011}a_{110}&+a_{010}a_{100}a_{011}a_{101})\\
+~4\,(a_{000}a_{011}a_{101}a_{110} &+ a_{001}a_{010}a_{100}a_{111}).
\end{split}
\end{array}
\end{equation}
The hyperdeterminant is invariant under $SL(2)_{A} \times SL(2)_{B} \times SL(2)_{C}$, with $a_{ABC}$ transforming as a $(\textbf{2},\textbf{2},\textbf{2})$, and under a discrete triality that interchanges $A$, $B$ and $C$. Another useful quantity is the local entropy $S_A$, which is a measure of how entangled $A$ is with the pair $BC$:
\begin{equation}
S_A=4\det\rho_A,
\end{equation}
where $\rho_A$ is the reduced density matrix
\begin{equation}
\rho_A=\Tr_{BC}|\Psi\rangle\langle\Psi|,
\end{equation}
and with similar formulae for $B$ and $C$.\par
\begin{table}
\caption{Three-qubit interpretation of the 8-charge $D=4$ black hole from four D3-branes wrapping around the lower four cycles of $T^6$ with wrapping numbers $N_0$, $N_1$, $N_2$, $N_3$ and then allowing $N_3$ to intersect at an angle $\theta$.}\label{tab:3QubitIntersect}
\begin{ruledtabular}
\begin{tabular}{*{11}{c}}
4 & 5 & & 6 & 7 & & 8 & 9 & Macrocharges & Microcharges & $|ABC\rangle$ \\
\hline
\textsf{x} & \textsf{o} & & \textsf{x} & \textsf{o} & & \textsf{x} & \textsf{o} & $p^0$  & 0                          & $|000\rangle$ \\
\textsf{o} & \textsf{x} & & \textsf{o} & \textsf{x} & & \textsf{x} & \textsf{o} & $q_1$  & 0                          & $|110\rangle$ \\
\textsf{o} & \textsf{x} & & \textsf{x} & \textsf{o} & & \textsf{o} & \textsf{x} & $q_2$  & $-N_3\sin\theta\cos\theta$ & $|101\rangle$ \\
\textsf{x} & \textsf{o} & & \textsf{o} & \textsf{x} & & \textsf{o} & \textsf{x} & $q_3$  & $N_3\sin\theta\cos\theta$  & $|011\rangle$ \\
\hline
\textsf{o} & \textsf{x} & & \textsf{o} & \textsf{x} & & \textsf{o} & \textsf{x} & $q_0$  & $N_0+N_3\sin^2\theta$      & $|111\rangle$ \\
\textsf{x} & \textsf{o} & & \textsf{x} & \textsf{o} & & \textsf{o} & \textsf{x} & $-p^1$ & $ -N_3\cos^2\theta$        & $|001\rangle$ \\
\textsf{x} & \textsf{o} & & \textsf{o} & \textsf{x} & & \textsf{x} & \textsf{o} & $-p^2$ & $-N_2$                     & $|010\rangle$ \\
\textsf{o} & \textsf{x} & & \textsf{x} & \textsf{o} & & \textsf{x} & \textsf{o} & $-p^3$ & $-N_1$                     & $|100\rangle$
\end{tabular}
\end{ruledtabular}
\end{table}
Two states of a composite quantum system are regarded as equivalent if they are related by a unitary transformation which factorises into separate transformations on the component parts, so-called \textit{local unitaries}. The Hilbert space decomposes into equivalence classes, or \textit{orbits} under the action of the group of local unitaries. For unnormalised three qubit states, the number of parameters \cite{Linden:1997qd} needed to describe inequivalent states or, what amounts to the same thing, the number of algebraically independent invariants \cite{Sudbery:2001} is given by the dimension of the space of orbits
\begin{equation}
\frac{\mathds{C}^2 \times \mathds{C}^2 \times \mathds{C}^2}{U(1) \times SU(2) \times SU(2) \times SU(2)},
\end{equation}
namely $16-10=6$. For subsequent comparison with the $STU$ black hole, however, we restrict our attention to states with \textit{real} coefficients $a_{ABC}$. In this case, one can show that there are five algebraically independent invariants: $\Det a,S_A,S_B,S_C$ and the norm $\langle\Psi|\Psi\rangle$, corresponding to the dimension of
\begin{equation}
\frac{\mathds{R}^2 \times \mathds{R}^2 \times \mathds{R}^2}{SO(2) \times SO(2) \times SO(2)},
\end{equation}
namely $8-3=5$. Hence, the most general real three-qubit state can be described by just five parameters \cite{Acin:2001}, conveniently taken as four real numbers $N_0,N_1,N_2,N_3$ and an angle $\theta$ \footnote{This is obtained from the canonical form for real states, Eq. (11) of \cite{Acin:2001}, by applying two different $SO(2)$ transformations on the second and third bits.}:
\begin{equation}\label{eq:General3QubitState}
\begin{array}{c}
|\Psi\rangle=-N_3\cos^2\theta|001\rangle-N_2|010\rangle \\
+~N_3\sin\theta\cos\theta|011\rangle-N_1|100\rangle  \\
-~N_3\sin\theta\cos\theta|101\rangle+(N_0+N_3\sin^2\theta)|111\rangle.
\end{array}
\end{equation}
With the more coarse-grained equivalence under $[SL(2)]^3$, one obtains the classification of three qubit entanglements shown in \autoref{tab:3QubitEntangClassif} \cite{Dur:2000}. Representatives from each class are: Class $A$-$B$-$C$ (product states)
\begin{equation}
N_0|111\rangle.
\end{equation}
Classes $A$-$BC$, (bipartite entanglement)
\begin{equation}
N_0|111\rangle-N_1|100\rangle,
\end{equation}
and similarly $B$-$CA$, $C$-$AB$. Class W (maximises bipartite entanglement)
\begin{equation}
-N_1|100\rangle- N_2|010\rangle-N_3|001\rangle.
\end{equation}
Class GHZ (genuine tripartite entanglement)
\begin{equation}\label{eq:ClassGHZState}
N_0|111\rangle-N_1|100\rangle- N_2|010\rangle-N_3|001\rangle.
\end{equation}
\par The $STU$ model \cite{Duff:1995sm} consists of $N=2$ supergravity coupled to three vector multiplets interacting through the special K\"{a}hler manifold $[SL(2)/SO(2)]^3$. A general static spherically symmetric black hole solution depends on 8 charges \cite{Duff:1995sm,Behrndt:1996hu} denoted $q_0$, $q_1$, $q_2$, $q_3$, $p^0$, $p^1$, $p^2$, $p^3$, but the generating solution depends on just $8-3=5$ parameters \cite{Cvetic:1995kv,Cvetic:1996zq}, after fixing the action of the isotropy subgroup $[SO(2)]^3$.  The solution can usefully be embedded in an $N=4$ supergravity model with symmetry $SL(2) \times SO(6,22)$, the low-energy limit of the heterotic string compactified on $T^6$, where the charges transform as a $(\textbf{2},\textbf{28})$, or else in $N=8$ supergravity with symmetry $E_{7(7)}$, the low-energy limit of the Type IIA or Type IIB strings, compactified on $T^6$ or M-theory on $T^7$, where the charges transform as a \textbf{56}. In all cases, remarkably, the same five parameters suffice to describe these 56-charge black holes \cite{Cvetic:1995kv,Cvetic:1996zq}.

By identifying the 8 charges with the 8 components of the three-qubit hypermatrix $a_{ABC}$, one finds \cite{Duff:2006uz} that the black hole entropy \cite{Behrndt:1996hu} is related to the 3-tangle as in \eqref{eq:Correspondence}. One also finds \cite{Kallosh:2006zs} in the $N=2$ theory that the three-qubit entanglement classification, discussed above, is matched by the black hole classification into small ($S=0$), with 1/2 of supersymmetry preserved, and large ($S\neq 0$),  with either 1/2 or 0. By embedding in the $N=8$  theory, we can in this Letter include the finer supersymmetry preserving distinctions \cite{Ferrara:1997ci} as in \autoref{tab:3QubitEntangClassif}.

There is, in fact, a quantum information theoretic interpretation of the 56 charge $N=8$ black hole in terms of a Hilbert space consisting of 7 copies of the three-qubit Hilbert space \cite{Duff:2006ue,Levay:2006pt}. It relies on the decomposition $E_{7(7)} \supset [SL(2)]^7$ and admits the interpretation, via the Fano plane, of a tripartite entanglement of seven qubits, with the entanglement measure given by Cartan's quartic $E_{7(7)}$ invariant.  Remarkably, however, because the generating solution depends on the same five parameters as the $STU$ model, its classification of states will exactly parallel that of the usual three qubits. Indeed, the Cartan invariant reduces to Cayley's hyperdeterminant \eqref{eq:Hyperdeterminant} in a canonical basis \cite{Kallosh:2006zs}.

Now we turn to the microscopic analysis. This is not unique since there are many ways of embedding the $STU$ model in string/M-theory, but a useful one from our point of view is that of four D3-branes of Type IIB wrapping the $(579),(568),(478),(469)$ cycles of $T^6$ with wrapping numbers $N_0,N_1,N_2,N_3$ and intersecting over a string \cite{Klebanov:1996mh}. The wrapped circles are denoted by crosses and the unwrapped circles by noughts as shown in \autoref{tab:3QubitIntersect}. This picture is consistent with the interpretation of the 4-charge black hole as a bound state at threshold of four 1-charge black holes \cite{Duff:1994jr,Duff:1996qp,Duff:1995sm}. The fifth parameter $\theta$ is obtained \cite{Balasubramanian:1997ak,Bertolini:2000ei} by allowing the $N_3$ brane to intersect at an angle which induces additional effective charges on the $(579),(569),(479)$ cycles. The microscopic calculation of the entropy consists of taking the logarithm of the number of microstates and yields the same result as the macroscopic one \cite{Bertolini:2000yaa}.

To make the black hole/qubit correspondence we associate the three $T^2$ with the $SL(2)_A \times SL(2)_B \times SL(2)_C$ of the three qubits Alice, Bob, and Charlie. The 8 different cycles then yield 8 different basis vectors $|ABC\rangle$ as in the last column of \autoref{tab:3QubitIntersect}, where $|0\rangle$ corresponds to \textsf{xo} and $|1\rangle$ to \textsf{ox}. We see immediately that we reproduce the five parameter three-qubit state $|\Psi\rangle$ of \eqref{eq:General3QubitState}. Note that the GHZ state of \eqref{eq:ClassGHZState} describes four D3-branes intersecting over a string. Performing a $T$-duality transformation, one obtains a Type IIA interpretation with zero D6 -branes, $N_0$ D0-branes, $N_1$, $N_2$, $N_3$ D4-branes plus effective D2-brane charges, where $|0\rangle$ now corresponds to \textsf{xx} and $|1\rangle$ to \textsf{oo}.

All this suggests that the analogy \cite{Duff:2007wa} between $D=5$ black holes and three-state systems (0 or 1 or 2), known as qutrits, should involve the choice of wrapping an M2-brane around one of three circles in $T^3$. This is indeed the case, with the number of qutrits being two.

The two qutrit system (where $A,B=0,1,2$) is described by the state
\begin{equation}
|\Psi\rangle=a_{AB}|AB\rangle,
\end{equation}
and the Hilbert space has dimension $3^2=9$. The bipartite entanglement of Alice and Bob is given by the 2-tangle
\begin{equation}
\tau_{AB}=27\det\rho_{A}=27\,|\det a_{AB}|^2,
\end{equation}
where $\rho_A$ is the reduced density matrix
\begin{equation}
\rho_A=\Tr_B|\Psi\rangle\langle\Psi|.
\end{equation}
The determinant is invariant under $SL(3)_A \times SL(3)_B$, with $a_{AB}$ transforming as a $(\textbf{3},\textbf{3})$, and under a discrete duality that interchanges $A$ and $B$.

Once again, for subsequent comparison with the $D=5$ black hole, we restrict our attention to unnormalised states with real coefficients $a_{AB}$. In this case, one can show \cite{Coffman:1999jd} that there are three algebraically independent invariants:  $\tau_{AB}$, $C_2$ (the principal minor of $\rho_A$) and the norm $\langle\Psi|\Psi\rangle$, corresponding to the dimension of the space of orbits
\begin{equation}
\frac{\mathds{R}^3 \times \mathds{R}^3}{SO(3) \times SO(3)},
\end{equation}
namely $9-6=3$. Hence, the most general two-qutrit state can be described by just three parameters, which may conveniently be taken to be three real numbers $N_0,N_1,N_2$:
\begin{equation}\label{eq:General2QutritState}
|\Psi\rangle = N_0|00\rangle+N_1|11\rangle+N_2|22\rangle.
\end{equation}
A classification of two-qutrit entanglements, depending on the rank of the density matrix, is given in \autoref{tab:2QutritEntangClassif}.\par
\begin{table}
\caption{Classification of two-qutrit entanglements and their corresponding $D=5$ black holes.}\label{tab:2QutritEntangClassif}
\begin{ruledtabular}
\begin{tabular}{crrcc}
Class       & $C_2$ & $\tau_{AB}$ & Black hole & SUSY \\
\hline
$A$-$B$     & 0     & 0           & small      & 1/2  \\
Rank 2 Bell & $>0$  & 0           & small      & 1/4  \\
Rank 3 Bell & $>0$  & $>0$        & large      & 1/8
\end{tabular}
\end{ruledtabular}
\end{table}
The 9-charge $N=2$, $D=5$ black hole may also be embedded in the $N=8$ theory in different ways. The most convenient microscopic description is that of three M2-branes \cite{Papadopoulos:1996uq,Klebanov:1996mh} wrapping the (58), (69), (710) cycles of the $T^6$ compactification of $D=11$ M-theory, with wrapping numbers $N_0,N_1,N_2$ as in \autoref{tab:2QutritIntersect}. To make the black hole/qutrit correspondence we associate the two $T^3$ with the $SL(3)_A \times SL(3)_B$ of the two qutrits Alice and Bob. The 9 different cycles then yield the 9 different basis vectors $|AB\rangle$ as in the last column of \autoref{tab:2QutritIntersect}, where $|0\rangle$ corresponds to \textsf{xoo}, $|1\rangle$ to \textsf{oxo}, and $|2\rangle$ to \textsf{oox}. We see immediately that we reproduce the three parameter two-qutrit state $|\Psi\rangle$ of \eqref{eq:General2QutritState}.

The black hole entropy, both macroscopic and microscopic, turns out to be given by the 2-tangle
\begin{equation}
S=2\pi\sqrt{|\det a_{AB}|},
\end{equation}
and the classification of the two-qutrit entanglements matches that of the black holes as in \autoref{tab:2QutritEntangClassif}.\par
\begin{table}
\caption{Two-qutrit interpretation of the 9-charge $D=5$ black hole from M2-branes in $D=11$ wrapping around the upper three cycles of $T^6$ with wrapping numbers $N_0,N_1,N_2$.}\label{tab:2QutritIntersect}
\begin{ruledtabular}
\begin{tabular}{*{11}{c}}
5 & 6 & 7 & & 8 & 9 & 10 & Macrocharges & Microcharges & $|AB\rangle$ \\
\hline
\textsf{x} & \textsf{o} & \textsf{o} & & \textsf{x} & \textsf{o} & \textsf{o} & $p^0$ & $N_0$ & $|00\rangle$ \\
\textsf{o} & \textsf{x} & \textsf{o} & & \textsf{o} & \textsf{x} & \textsf{o} & $p^1$ & $N_1$ & $|11\rangle$ \\
\textsf{o} & \textsf{o} & \textsf{x} & & \textsf{o} & \textsf{o} & \textsf{x} & $p^2$ & $N_2$ & $|22\rangle$ \\
\hline
\textsf{x} & \textsf{o} & \textsf{o} & & \textsf{o} & \textsf{x} & \textsf{o} & $p^3$ & $0$   & $|01\rangle$ \\
\textsf{o} & \textsf{x} & \textsf{o} & & \textsf{o} & \textsf{o} & \textsf{x} & $p^4$ & $0$   & $|12\rangle$ \\
\textsf{o} & \textsf{o} & \textsf{x} & & \textsf{x} & \textsf{o} & \textsf{o} & $p^5$ & $0$   & $|20\rangle$ \\
\hline
\textsf{x} & \textsf{o} & \textsf{o} & & \textsf{o} & \textsf{o} & \textsf{x} & $p^6$ & $0$   & $|02\rangle$ \\
\textsf{o} & \textsf{x} & \textsf{o} & & \textsf{x} & \textsf{o} & \textsf{o} & $p^7$ & $0$   & $|10\rangle$ \\
\textsf{o} & \textsf{o} & \textsf{x} & & \textsf{o} & \textsf{x} & \textsf{o} & $p^8$ & $0$   & $|21\rangle$
\end{tabular}
\end{ruledtabular}
\end{table}
There is, in fact, a quantum information theoretic interpretation of the 27 charge $N=8,D=5$ black hole in terms of a Hilbert space consisting of three copies of the two-qutrit Hilbert space \cite{Duff:2007wa}. It relies on the decomposition $E_{6(6)} \supset [SL(3)]^3$ and admits the interpretation of a bipartite entanglement of three qutrits, with the entanglement measure given by Cartan's cubic $E_{6(6)}$ invariant. Once again, however, because the generating solution depends on the same three parameters as the 9-charge model, its classification of states will exactly parallel that of the usual two qutrits. Indeed, the Cartan invariant reduces to $\det a_{AB}$ in a canonical basis \cite{Ferrara:1997ci}.

In conclusion, our Type IIB microscopic analysis of the black hole has provided an explanation for the appearance of the qubit two-valuedness (0 or 1) that was lacking in the previous treatments: the brane can wrap one circle or the other in each $T^2$.  The number of qubits is three because of the six extra dimensions of string theory.  Moreover, the five parameters of the real three-qubit state are seen to correspond to four D3-branes intersecting at an angle. Similar results hold for the two-qutrit system. It would be interesting to see whether we can now find an underlying physical justification for \eqref{eq:Correspondence}.

\begin{acknowledgments}
We thank Dan Waldram for interesting discussions and Iosif Bena for useful correspondence. H.E. would like to thank the theoretical physics group at Imperial College for their warm hospitality. This research is supported in part by STFC under rolling Grant No. PP/D0744X/1.
\end{acknowledgments}

\end{document}